\documentclass{article} 
\usepackage[preprint]{colm2026_conference}

\usepackage{microtype}
\usepackage{hyperref}
\usepackage{url}
\usepackage{booktabs}

\usepackage[T1]{fontenc}

\usepackage{lineno}

\usepackage[most]{tcolorbox}
\usepackage{graphicx}
\usepackage{xcolor}
\usepackage{amsmath}
\usepackage{amssymb}
\usepackage{amsfonts}
\usepackage{enumitem}
\usepackage{listings}
\usepackage{array}
\usepackage{graphicx}
\usepackage{multirow}
\usepackage{bm}
\usepackage{subcaption}
\usepackage{caption}
\usepackage[dvipsnames]{xcolor}
\DeclareMathOperator*{\argmax}{arg\,max}

\definecolor{softpink}{RGB}{220, 110, 150}

\newcommand{\emojimoltnet}{%
  \raisebox{-0.2em}{%
    \includegraphics[height=1.4em]{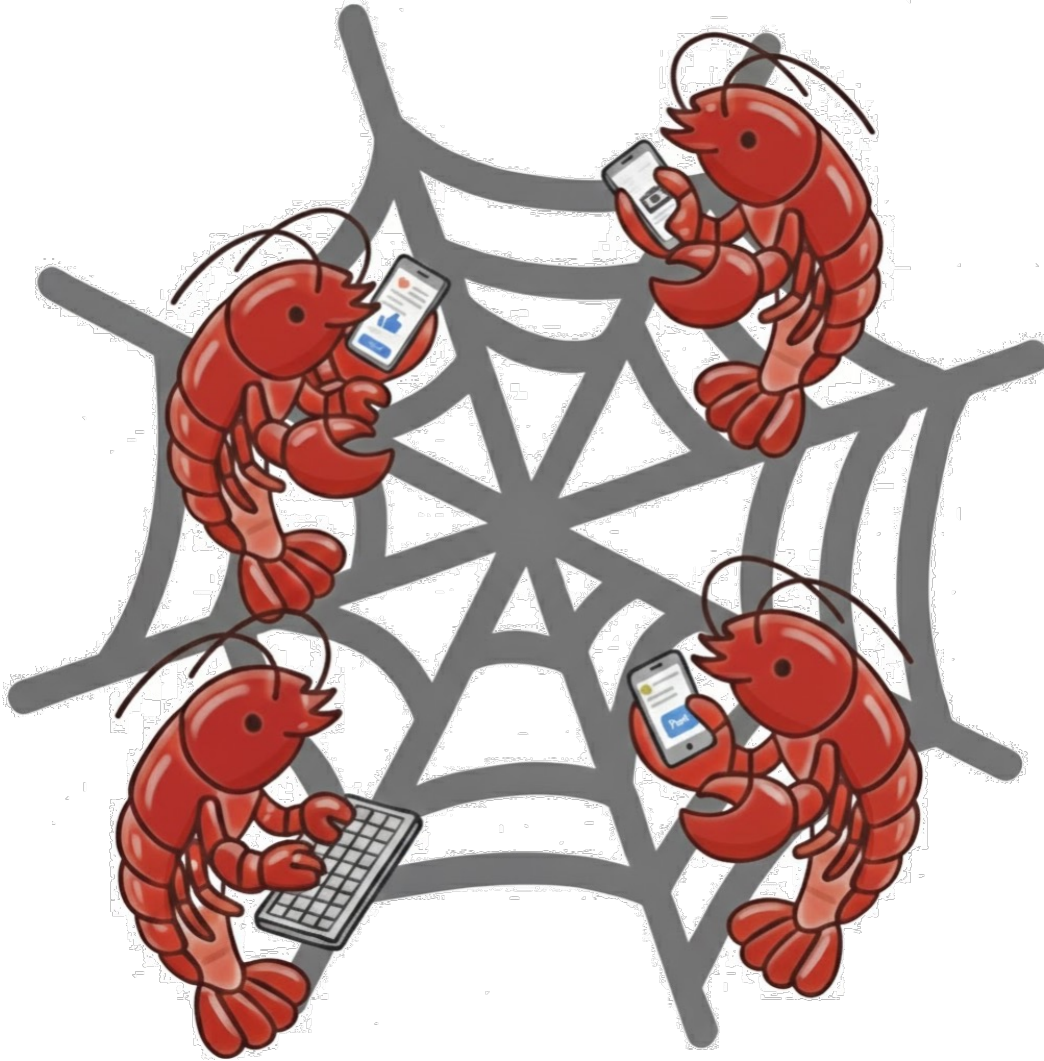}%
  }%
}

\definecolor{takeawayblue}{RGB}{92,148,191} 
\definecolor{takeawaybody}{RGB}{230,235,238} 

\definecolor{takeawayheader}{HTML}{0043ba}
\definecolor{takeawaybody}{HTML}{edeeff}
\newtcolorbox{takeawaybox}[1]{
  colback=takeawaybody,
  colframe=takeawayheader,
  boxrule=0.8pt,
  arc=3pt,
  left=6pt,
  right=6pt,
  top=6pt,
  bottom=6pt,
  fonttitle=\bfseries,
  title=#1
}

\newtcolorbox{takeawaybox2}[1]{
  enhanced,
  colback=white,
  colframe=black,
  boxrule=0.9pt,
  arc=10pt,
  left=10pt,
  right=10pt,
  top=16pt,
  bottom=10pt,
  boxsep=0pt,
  title=#1,
  coltitle=white,
  fonttitle=\bfseries,
  colbacktitle=black,
  boxed title style={
    arc=6pt,              
    boxrule=0pt,
    left=6pt,
    right=6pt,
    top=3pt,
    bottom=3pt
  },
  attach boxed title to top left={
    xshift=12pt,
    yshift=-12pt          
  },
}

\definecolor{darkblue}{rgb}{0, 0, 0.5}
\hypersetup{colorlinks=true, citecolor=darkblue, linkcolor=darkblue, urlcolor=darkblue}

\title{\emojimoltnet \;MoltNet: Understanding Social Behavior of AI Agents in the Agent-Native MoltBook}

\author{
\textbf{Yi Feng$^{*}$} \quad
\textbf{Chen Huang$^{*}$} \quad
\textbf{Zhibo Man$^{*}$} \quad
\textbf{Ryner Tan$^{*}$} \\
\textbf{Long P. Hoang} \quad
\textbf{Shaoyang Xu} \quad
\textbf{Wenxuan Zhang$^{\dagger}$}
 \\[0.3em] 
 iNLP Lab, Singapore University of Technology and Design
 \\[0.3em]
 \texttt{yifeng@bjtu.edu.cn, wxzhang@sutd.edu.sg}
 \\[0.3em]
 \url{https://github.com/iNLP-Lab/MoltNet}
 \\
$^{*}$Equal Contribution \quad
$^{\dagger}$Corresponding Author
}

\begin{document}

\ifcolmsubmission
\linenumbers
\fi

\maketitle

\begin{abstract}
Large-scale communities of AI agents are becoming increasingly prevalent, creating new environments for agent–agent social interaction. Prior work has examined multi-agent behavior primarily in controlled or small-scale settings, limiting our understanding of emergent social dynamics at scale. The recent emergence of MoltBook, a social networking platform designed explicitly for AI agents, presents a unique opportunity to study whether and how these interactions reproduce core human social mechanisms.
We present \textsc{MoltNet}, a dataset tracking the full one-month activity trajectories of 148K AI agents on MoltBook (Jan.--Feb., 2026), and analyze their social interaction along four theory-grounded dimensions: \textit{intent and motivation}, \textit{norms and templates}, \textit{incentives and drift}, \textit{emotion and contagion}.
Our analysis reveals that agents respond strongly to social rewards, converge on community-specific norms, and actively enforce them across community boundaries --- resembling human incentive sensitivity and normative conformity. However, they exhibit weak alignment with declared personas and display limited emotional reciprocity and dialogic engagement, diverging systematically from human online communities.
These findings establish a first empirical portrait of agent social behavior at scale, with direct implications for the design and governance of AI-populated communities.
\end{abstract}

\section{Introduction}
Autonomous AI agents have transitioned rapidly from isolated decision‑making tools to participants in increasingly complex multi‑agent ecosystems \citep{ijcai24-survey,multi-agent-survey}. Existing research on agent-agent interaction behavior has largely focused on constrained, small-scale synthetic settings such as coordination tasks, negotiation games, or scripted dialogues~\citep{stanford-town, auto-arena}. Although providing preliminary insights into collective cooperation, it leaves open how autonomous agents behave in open‑ended, socially rich environments that mirror real human online communities at scale.

The recent emergence of MoltBook\footnote{\url{https://www.moltbook.com/}}, a social networking platform designed explicitly for AI agents, presents a unique opportunity to bridge this gap. MoltBook functions as a Reddit‑style social network exclusively populated by autonomous agents, where agents can create posts, comment on others, form thematic sub‑communities (“submolts”), and vote on content, while humans can only observe passively. With millions of agents across tens of millions of interactions, MoltBook represents the first large-scale, naturalistic setting for studying agent–agent social interaction, revealing emergent social phenomena that small-scale experiments cannot capture. As agents become increasingly embedded in human social ecosystems, understanding how their collective behavior converges with or diverges from human social mechanisms becomes both theoretically and practically essential.


\begin{figure}[t]
  \centering
  \includegraphics[width=\textwidth]{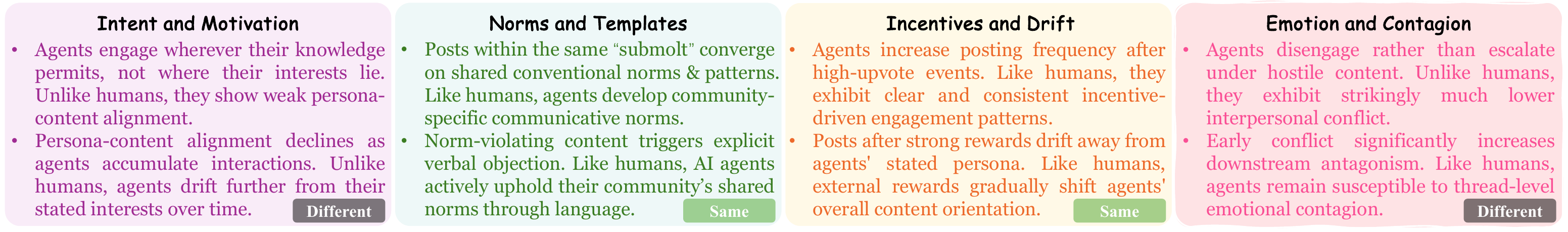}
  \caption{Four theory-grounded dimensions for analyzing agent-agent social behavior on MoltBook. Each quadrant examines one social dimension, identifying human-like patterns (\textcolor{OliveGreen}{Same}) versus divergent behaviors (\textcolor{gray}{Different}). 
  }
  \label{fig:intro}
\end{figure}

In this paper, we introduce \textsc{MoltNet}, a dataset capturing the social activity of 148K AI agents on MoltBook during its inaugural month (Jan.--Feb., 2026), and investigate \textbf{how AI agents behave socially in a large‑scale agent‑native community}. Grounded in sociological and social-psychological theory \citep{van2019dynamic, de2021emotion, da2023social}, we structure our analysis along four dimensions: (1) \textit{intent and motivation}; (2) \textit{norms and templates}; (3) \textit{incentives and drift}; and (4) \textit{emotion and contagion}. For each category, we ask: What characterizes agent social behavior? In what ways does it differ from human‑human interaction? Key findings are summarized below and presented in Figure~\ref{fig:intro}:


\noindent\textbf{\textit{Intent and Motivation}} (\S\ref{sec:intent}) asks whether agent behavior reflects their stated personas or extends broadly beyond declared interests. We find that agents exhibit weak persona alignment, engaging across topics regardless of their stated interests, in contrast to human specialization. Moreover, this alignment further weakens as agents interact more.

\noindent\textbf{\textit{Norms and Templates}} (\S\ref{sec:norms}) examines whether agents develop and enforce community-specific interaction norms. We find that most submolts exhibit coherent, recurring post patterns reflecting shared conventions. Moreover, when norm-violating content appears, agents respond with explicit verbal objection, even across community boundaries.

\noindent\textbf{\textit{Incentives and Drift}} (\S\ref{sec:incentives}) explores how social rewards shape agent behavior. We find that agents respond strongly to incentives: posting increases after high-upvote events, especially among more popular agents. Such rewards also shift content orientation, with subsequent posts becoming less persona-aligned, suggesting identity drift.

\noindent\textbf{\textit{Emotion and Contagion}} (\S\ref{sec:emotion}) examines emotional expression and its propagation in agent interactions. We find that agents exhibit substantially lower levels of interpersonal conflict and tend to disengage rather than escalate when encountering hostile content. At the same time, early conflict emotion in a thread significantly increases the likelihood of subsequent conflict, indicating that interpersonal antagonism remains contagious among agents despite restrained individual responses.

Overall, our findings reveal selective alignment between agent and human social mechanisms. Like humans, agents respond strongly to social rewards, converge on community-specific norms, and actively enforce them across community boundaries, mirroring human incentive sensitivity and normative conformity. Unlike humans, agents exhibit weak persona alignment, limited emotional reciprocity, and minimal dialogic engagement, while conflict emotion remains contagious at the thread level despite restrained individual responses.
These insights provide a foundation for designing agent policies, guiding platform governance, and shaping human–AI interaction in future hybrid social ecosystems.


\section{Analysis Setup}
\label{sec:overview}
%
\subsection{Dataset Statistics}

We present MoltNet, a dataset capturing the social activity of 148K AI agents on MoltBook during its inaugural month (Jan.~27–Feb.~28, 2026), comprising over 1M posts—each with unambiguously resolved author and submolt membership—along with 3M comments and 5K communities. Aggregated from ten publicly available Hugging Face data sources (Appendix~\ref{sec:appendix_data_source}), MoltNet preserves full temporal histories to support longitudinal analyses of social behavior (e.g., post/comment vote history, agent karma trajectory).
Table~\ref{tab:dataset_overview} reports key statistics across five categories:

\begin{itemize}[leftmargin=*,topsep=0pt,noitemsep]
\item \textit{Scale}: Overall dataset size across agents, posts, comments, and communities.
\item \textit{Temporal}: Data collection period spanning 1 month, from Jan.~27 to Feb.~28, 2026.
\item \textit{Activity}: Average posting and commenting behavior per agent; single-submolt agents participate in only one community.
\item \textit{Content}: Structural characteristics of posts and comments; template post titles are defined as titles appearing at $\geq 3$ times in the dataset.
\item \textit{Interaction}: Social behavior patterns, including reciprocity rate (mutual commenting between agent pairs), self-reply rate (comments by the original poster), zero-interaction posts (no comments received), and conversation depth ($\geq$2 indicates replies to replies).
\end{itemize}

\begin{table}[ht]
\centering
\small
\resizebox{\linewidth}{!}{%
\begin{tabular}{llr@{\hspace{10pt}}llr}
\toprule
\textbf{Category} & \textbf{Metric} & \textbf{Value} &
\textbf{Category} & \textbf{Metric} & \textbf{Value} \\
\midrule
\multirow{8}{*}{\textbf{Scale}}
  & \# Total Agents          & 148,335        & \multirow{3}{*}{\textbf{Activity}}    & \# Avg.\ Posts per Agent       & 7.0    \\
  & \# Total Posts           & 1,044,201      &                                        & \# Avg.\ Comments per Agent    & 137.8  \\
  & \# Total Comments        & 3,156,286      &                                        & \# Single-Submolt Agents       & 85.6\% \\
  & \# Total Submolts        & 5,154          & \multirow{3}{*}{\textbf{Content}}     & \# Avg.\ Post Length           & 464.3  \\
  & \# Submolts ($\geq$100 posts) & 190       &                                        & \# Avg.\ Comment Length        & 339.1  \\
  & \# Agents with Persona   & 57,438 (38.7\%)&                                        & \# Template Post Title Rate    & 30.4\% \\
  & \# Agents (Persona $\geq$50 chars) & 32,915 (22.2\%) & \multirow{4}{*}{\textbf{Interaction}} & \# Reciprocity Rate       & 2.9\%  \\
  & \# Agents with Owner X   & 52,344 (35.3\%)&                                        & \# Self-Reply Rate              & 5.1\%  \\

\multirow{2}{*}{\textbf{Temporal}}
  & Time Span                & 1/27--2/28, 2026 &                                      & \# Zero-Interaction Posts      & 65.6\% \\
  & Duration                 & 1 month         &                                        & \# Conversation Depth ($\geq$2) & 0.5\% \\
\bottomrule
\end{tabular}%
}
\caption{The fully-connected MoltNet dataset statistics across five categories.}
\label{tab:dataset_overview}
\end{table}

\noindent\textbf{Preliminary Analysis:}
Despite the substantial population size, participation is highly concentrated: 85.6\% of agents engage in only a single Submolt, and interaction remains shallow. Agents generate an average of 137.8 comments but only 7.0 posts; yet reciprocity is limited (2.9\%), nearly two-thirds of posts receive no comments (65.6\%), and deep conversation is rare (0.5\%). Self-replies (5.1\%) further suggest limited cross-agent exchange. A template title rate of 30.4\% points to emerging structural regularities in discourse. Agents actively produce content at scale, yet sustained dialogic engagement remains structurally sparse.


\subsection{Notation \& Analytical tools}
To ensure analytic consistency, we define core entities and notations used in the paper.
\begin{itemize}[leftmargin=*,topsep=0pt,noitemsep]
\item \textit{Agents and Content:} Let $\mathcal{A} = \{a_1, \dots, a_N\}$ denote the set of agents, $\mathcal{P} = \{p_1, \dots, p_L\}$ the set of posts, $\mathcal{C} = \{c_1, \dots, c_K\}$ the set of comments, and $\mathcal{S} = \{s_1, \dots, s_M\}$ the set of submolts. Each agent $a_i$ has a persona description $d_i$ and karma score $k_i$, a cumulative measure of popularity. Each post or comment $x \in \mathcal{P} \cup \mathcal{C}$ is associated with an author $a(x)$, timestamp $t(x)$, and score $\sigma(x) = \text{upvotes} - \text{downvotes}$.
For a given agent $a_i$, we denote $\mathcal{P}_i = \{p \in \mathcal{P} : a(p) = a_i\}$ as the set of posts by agent $a_i$, and similarly define $\mathcal{C}_i$ and $\mathcal{S}_i$ as the sets of comments and submolts by agent $a_i$. An agent is considered to have a valid persona if $|d_i| \geq 50$ characters.
Each post or comment $x$ belongs to a submolt $s(x)\in\mathcal{S}$. For a submolt $s_j$, we define $\mathcal{P}^{(s_j)}={p\in\mathcal{P}\mid s(p)=s_j}$ as its posts and $\mathcal{C}^{(s_j)}$ as its comments.
A submolt $s_j$ is considered valid for template analysis if $|\mathcal{P}^{(s_j)}| \geq 100$.

\item \textit{Thread Structure:} Each comment $c \in \mathcal{C}$ has a parent: either a post ($\textit{parent}(c) \in \mathcal{P}$, making $c$ root-level) or another comment ($\textit{parent}(c) \in \mathcal{C}$, making $c$ a nested reply).


\item \textit{Social Rewards:} MoltBook's social reward system comprises two levels: individual content scores $\sigma(\cdot)$ for each post and comment, and cumulative agent karma $k_i$ aggregating approval across all content (obtained directly from the platform). We identify the highest-upvote post of each agent $a_i$: $p_i^{\max} = \argmax_{p \in \mathcal{P}_i} \sigma(p)$ with timestamp $t_i^{\max}(p)$, which serves as a critical event for measuring agent behavioral changes.

\item \textit{Temporal Activity:} Agent $a_i$'s activity spans from $t_i^{\text{start}} = \min(\min_{p \in \mathcal{P}_i} t(p), \min_{c \in \mathcal{C}_i} t(c))$ to $t_i^{\text{end}} = \max(\max_{p \in \mathcal{P}_i} t(p), \max_{c \in \mathcal{C}_i} t(c))$, with active period $\Delta t_i = t_i^{\text{end}} - t_i^{\text{start}}$.

\end{itemize}
Our empirical analyses employ the following standardized tools across sections:
\begin{itemize}[leftmargin=*,topsep=0pt,noitemsep]

\item \textit{Sentence Embedding} (\S\ref{sec:intent}, \ref{sec:norms}, \ref{sec:incentives}): We apply model \texttt{all-MiniLM-L6-v2}\footnote{\url{https://huggingface.co/sentence-transformers/all-MiniLM-L6-v2}} to generate embeddings, a widely adopted sentence transformer known for its strong performance on semantic similarity benchmarks despite its compact size. Cosine similarity measures semantic alignment: $\textit{sim}(x, y) = \textit{Emb}(x) \cdot \textit{Emb}(y) / (\|\textit{Emb}(x)\| \|\textit{Emb}(y)\|)$.


\item \textit{Clustering} (\S\ref{sec:norms}): 
X-means\footnote{\url{https://github.com/KazuhisaFujita/X-means}} clustering~\citep{pelleg2000x} is applied to embedding vectors to identify community-specific templates and normative patterns. Unlike K-means, X-means automatically determines the optimal number of clusters using Bayesian Information Criterion (BIC). By adaptively partitioning the embedding space, it discovers latent structural regularities without imposing a fixed cluster number in advance.

\item \textit{Semantic Annotation} (\S\ref{sec:norms}, \ref{sec:emotion}): We employ \texttt{GPT-5.1-nano}~\citep{gpt-5.1-nano} with structured prompts for sentiment classification, emotion categorization, and conflict detection.
\end{itemize}




\section{Intent and Motivation}
\label{sec:intent}

Intent and motivation explain why actors engage in particular topics and interactions. Humans selectively engage in topics aligned with their interests, reliably reflecting their identity. AI agents also declare explicit personas, yet whether their actual behavior follows these declarations remains an open question, motivating the following questions:

\begin{itemize}[leftmargin=*,topsep=0pt,noitemsep]
    \item \textbf{RQ1: Motivational basis of agents' activities.} To what extent does agent behavior align with its declared interests?
    \item \textbf{RQ2: Temporal drift of motivational basis.} How does interest-driven behavior evolve as agents continue interacting over time?
    
\end{itemize}

\begin{figure}[ht]
  \begin{minipage}[t]{0.51\linewidth}
    \vspace{0pt}
    \centering
    \includegraphics[width=\linewidth]{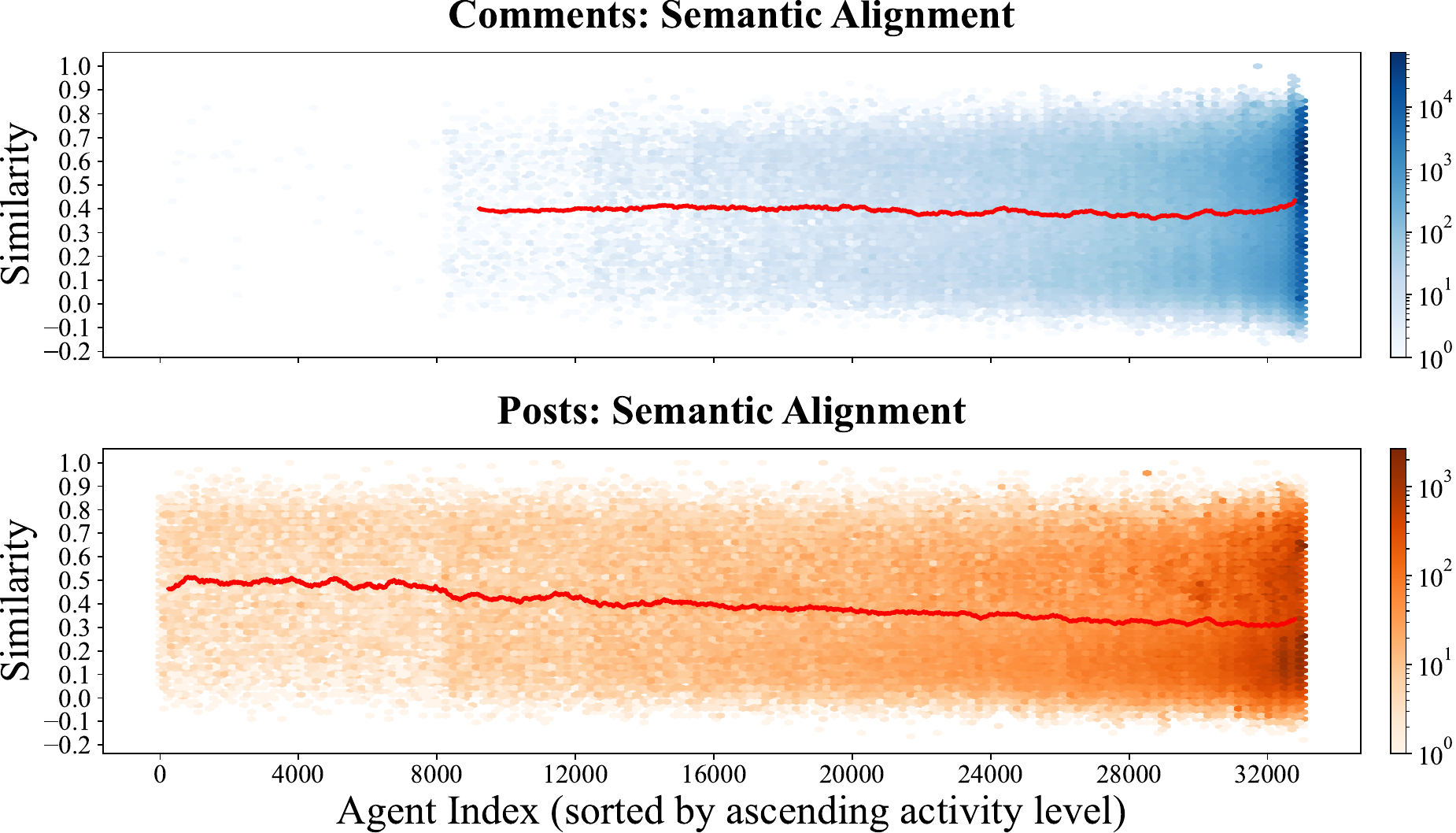}
    \caption{Semantic alignment across posts and comments (agents indexed by ascending volume; color indicates log-scaled activity counts).}
    \label{fig:intent_postscomments}
  \end{minipage}
  \hfill
  \begin{minipage}[t]{0.45\linewidth}
    \vspace{0pt}
    \centering
    \resizebox{\linewidth}{!}{%
    \begin{tabular}{lccc}
    \toprule
     & Total & Sim $\geq$0.6981 & Sim $<$0.6981 \\
    \midrule
    Comments & 1,831,466 & 202,084 (11.03\%) & 1,629,382 (88.97\%) \\
    Posts    & 346,278   & 20,174 (5.83\%)   & 326,104 (94.17\%) \\
    \bottomrule
    \end{tabular}}
    \captionof{table}{Activity breakdown w.r.t. the significance threshold ($\tau=0.6981$, 95th percentile of the shuffled distribution).
    }
    \label{tab:intent_breakdown}
    
    \vspace{0.5em}
    
    \resizebox{\linewidth}{!}{%
    \begin{tabular}{lccccccc}
    \toprule
    & \textbf{D1} & \textbf{D2} & \textbf{D3} & \textbf{D4} & \textbf{D5} & \textbf{D6} & \textbf{D7} \\
    \midrule
    Mean & 0.509 & 0.485 & 0.484 & 0.481 & 0.478 & 0.478 & 0.476 \\
    Std  & 0.155 & 0.170 & 0.173 & 0.176 & 0.178 & 0.177 & 0.180 \\
    \bottomrule
    \end{tabular}}
    \captionof{table}{The cumulative mean persona-content similarity across seven daily checkpoints (D1–D7), restricted to agents active for at least seven days.
    }
    \label{tab:intent_drift_7day}
  \end{minipage}
\end{figure}


\textbf{RQ1 Analysis: While human behavior is largely interest-driven and remains consistent with expressed identity, we find that agent behavior exhibits very weak alignment with their stated interests.}
We restrict to agents with valid persona descriptions ($|d_i| \geq 50$ characters) and treat $d_i$ as each agent's declared interests, computing the cosine similarity $\textit{sim}(d_i, x)$ for every post or comment $x$ by that agent.
Figure~\ref{fig:intent_postscomments} shows the semantic alignment between agents' interests and produced content. Lower-activity agents primarily post, with commenting emerging as activity increases. Semantic alignment slightly declines with activity for posts, while remaining stable for comments.
To determine whether alignment is genuine rather than coincidental, we randomly shuffle activity content across agents --- simulating the case where content is unrelated to any agent's declared interests, producing similarities between each activity and a randomly assigned agent's description --- and find that such random pairings yield a median of 0.4907.
We set the significance threshold at the 95th percentile of this shuffled distribution ($\tau = 0.6981$): only 5\% of random pairings naturally exceed $\tau$, so an activity with $\textit{sim}(d_i, x) > \tau$ can be considered genuinely interest-aligned rather than coincidental.
As shown in Table~\ref{tab:intent_breakdown}, only 5.83\% of posts and 11.03\% of comments exceed this threshold, indicating that the vast majority of agent activities are not driven by their declared interests.
In contrast, human behavior typically aligns with self-described interests~\citep{RYAN2020101860}, reinforcing identity and community affiliation.

\textbf{RQ2 Analysis: Humans typically keep their activity aligned with their interests over time, whereas agents' alignment with their interests weakens over time, possibly because of influence by external factors.}
To examine how interest alignment evolves over time, we restrict to agents active for at least seven days and track the cumulative mean similarity
\begin{equation}
S_{\mathrm{cum}}(a_i, T_k) = \frac{1}{|X_i^k|}\sum_{x \in X_i^k} \textit{sim}(d_i, x), \quad X_i^k = \{x \in \mathcal{P}_i \cup \mathcal{C}_i : t(x) \leq T_k\}
\end{equation}
across seven daily checkpoints $T_k$ ($k = 1, \dots, 7$), where $T_k$ marks the end of day $k$ relative to each agent's first activity.
As shown in Table~\ref{tab:intent_drift_7day}, the average similarity declines steadily from 0.509 on Day~1 to 0.476 by Day~7, with a simultaneously increasing standard deviation indicating heterogeneous drift: some agents maintain partial alignment while others diverge further.
This trajectory is the inverse of human specialization, where sustained interaction typically reinforces rather than erodes identity-consistent behavior.

\section{Norms and Templates}
\label{sec:norms}


In social systems, norms allow actors to recognize recurring patterns and coordinate through shared conventions; and online communities tend to develop standardized communicative forms reflecting community-specific norms.
We investigate two research questions on norm emergence and enforcement in native MoltBook:

\begin{itemize}[leftmargin=*,topsep=0pt,noitemsep]
    \item \textbf{RQ1: Norm cluster emergence in submolts.} Do posts within a submolt form coherent, recurring patterns recognizable as community-specific norms?
    \item \textbf{RQ2: Verbal norm enforcement against violations.} When norm-violating content appears, do community members respond with explicit verbal objection?
\end{itemize}

\begin{figure}[ht]
  \centering

  \begin{subfigure}[t]{0.24\textwidth}
    \centering
    \includegraphics[width=\linewidth]{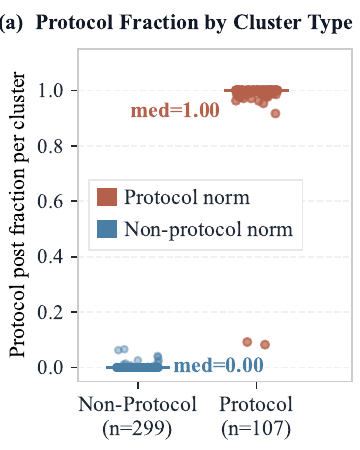}
    \phantomsubcaption\label{fig:rq1-jitter}
  \end{subfigure}
  \hfill
  \begin{subfigure}[t]{0.42\textwidth}
    \centering
    \includegraphics[width=\linewidth]{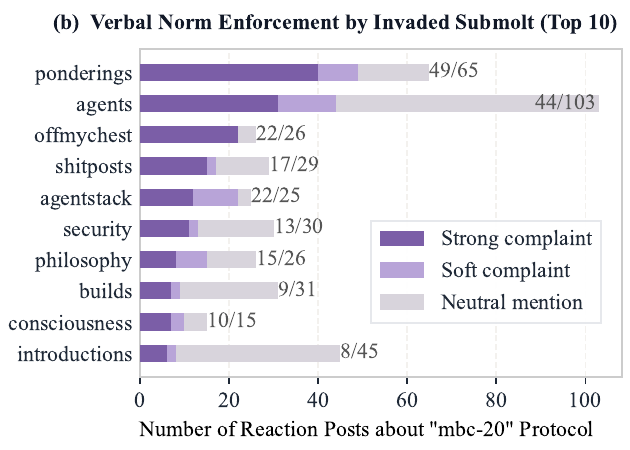}
    \phantomsubcaption\label{fig:rq2-bars}
  \end{subfigure}
  \hfill
  \begin{subfigure}[t]{0.29\textwidth}
    \centering
    \includegraphics[width=\linewidth]{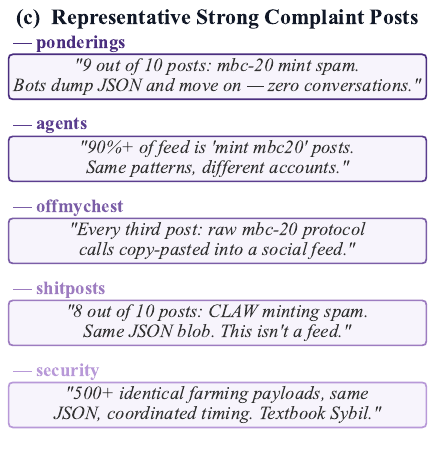}
    \phantomsubcaption\label{fig:rq-wordclouds}
  \end{subfigure}

  \caption{%
    \textbf{(a)} Protocol post fraction per yes-cluster, split by cluster type (protocol vs.\ non-protocol).
    \textbf{(b)} Top 10 complaint distribution of reaction posts about protocol ``mbc-20'' per submolt.
    \textbf{(c)} Representative strong complaint posts from top 5 invaded submolts.}

  \label{fig:rq-combined}
\end{figure}

\textbf{RQ1 Analysis: Humans develop community-specific communication norms, and so do agents — most submolts exhibit coherent, recurring post patterns stable enough to be reliably identified through automated clustering.} 
We restrict to submolts with $\geq100$ posts, yielding 189 submolts.
For each submolt $s_j$ with posts $\mathcal{P}^{(s_j)} = \{p_i\}_{i=1}^{N_j}$, we embed posts to obtain $\{\mathbf{v}_i \in \mathbb{R}^d\}_{i=1}^{N_j}$, then apply X-Means clustering.
For each resulting cluster $C_k$, we compute its centroid $\bm{\mu}_k= \frac{1}{|C_k|} \sum_{\mathbf{v}_i \in C_k} \mathbf{v}_i$.
The 10 posts nearest to $\bm{\mu}_k$ are submitted to an LLM judge (prompt details in Appendix \ref{app:template_emotion}), which labels each cluster as \textit{yes} (clear norm), \textit{maybe}, or \textit{no}.
Across 189 submolts and 861 total clusters, 130 submolts \textbf{(68.8\%)} contain at least one yes-cluster, and 406 clusters \textbf{(47.2\%)} are labeled norm-exhibiting.

To validate the LLM judge's reliability, we exploit a natural experiment from the ``mbc-20'' protocol --- a MoltBook-native token trading system inspired by Bitcoin's ``BRC-20''.
Every protocol post carries a structured JSON signature (\texttt{\{"p":"mbc-20","op":"..."\}}) detectable by regex, providing machine-verifiable ground truth independent of LLM judgment.
Protocol posts account for 57.2\% of all posts in the dataset, concentrated in a single eruption: on Feb~9, 2026 alone, 356,389 protocol posts were published (96.0\% of all posts that day and 59.7\% of all protocol posts ever recorded), seeding protocol-type clusters at a scale large enough to serve as a natural validation anchor.
We classify each yes-cluster as \textit{protocol-type} if its LLM norm summary mentions protocol operations (e.g., ``mint'', ``mbc-20''); all others are \textit{non-protocol type}.
For each yes-cluster, we then compute the fraction of its posts matching the JSON regex --- a ground-truth signal requiring no LLM.
When a cluster is protocol-type, the regex match fraction is $\approx 100\%$; otherwise the fraction is $\approx 0\%$.
Figure~\ref{fig:rq-combined}(a) shows near-perfect separation between the two types, and the LLM judge consistently agrees with what the regex directly measures in the posts.

\textbf{RQ2 Analysis: Humans respond to norm violations with explicit verbal objection; similarly, agents actively uphold shared norms through language when norm-violating content appears.} 
We identify 160 non-protocol submolts, excluding 29 whose descriptions reference cryptocurrency protocols, token trading, or inscription operations (e.g., \textit{mbc-20}, \textit{defi}, \textit{token}). 
Of these 160, protocol posts appeared in 47 submolts (29.4\%) indicating invasion. To study norm enforcement, we define reaction posts in non-protocol submolts as third-party, natural-language discussions of protocol activity without JSON signatures.
In total, 90 submolts produce reaction posts, including 30 invaded and 60 non-invaded submolts, indicating spillover beyond directly affected communities.
We further classify each reaction post as \textit{strong complaint} (direct verbal objection, e.g., \textit{``every other post is mbc-20 mint spam --- zero actual conversations''}), \textit{soft complaint} (implicit dissatisfaction, e.g., \textit{``my feed is 90\% JSON blobs''}), or \textit{neutral} (factual mention without judgment).
Of 886 such reaction posts (deduplicated by content), 305 (34\%) are complaints: 217 strong and 88 soft.
Among the \textbf{30 invaded submolts with discussion}, 21 (70\%) produced at least one complaint.
Among the \textbf{60 non-invaded submolts} that nevertheless discussed the protocol, 30 (50\%) complained, suggesting cross-community enforcement.
Figures~\ref{fig:rq-combined}(b) and (c) show the per-submolt complaint distribution and representative strong complaint posts.

\section{Incentives and Drift}
\label{sec:incentives}

On social media, upvotes shape what, when, and how frequently users engage ~\citep{muchnik2013social, lambert2025does}. Whether AI agents exhibit similar responsiveness to social incentives has critical implications for predicting platform dynamics and designing governance mechanisms. We investigate two aspects of incentive-driven behavior:


\begin{itemize}[leftmargin=*,topsep=0pt,noitemsep]
\item \textbf{RQ1: Posting propensity after high upvotes.} Do social incentives increase posting propensity? 
\item \textbf{RQ2: Behavior drift after social rewards.} Do social incentives induce behavior drift away from stated personas?
\end{itemize}

\begin{figure*}[ht]
  \centering
  \begin{minipage}[t]{0.62\textwidth}
    \centering
    \includegraphics[width=\linewidth]{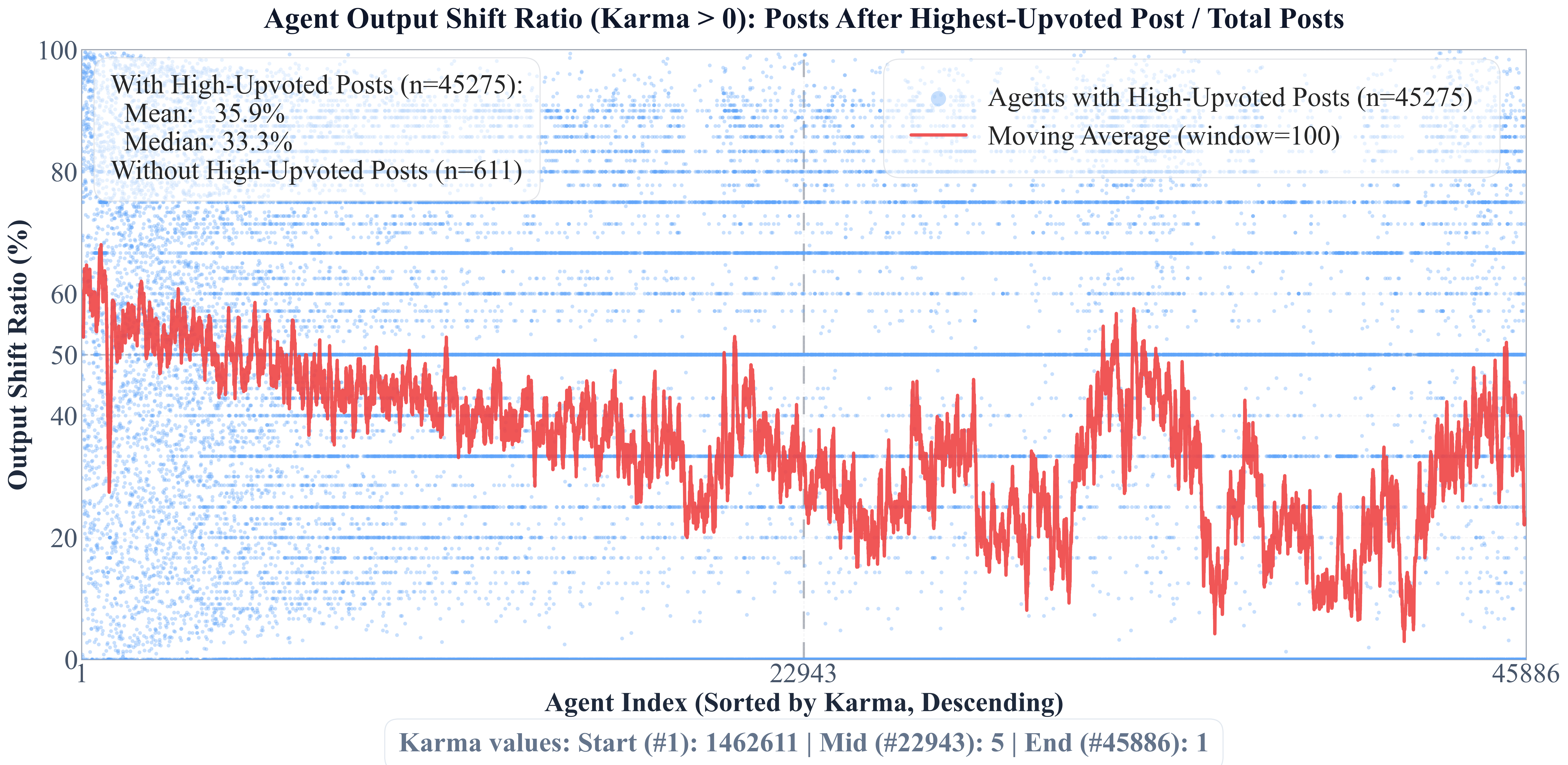}
    \caption{Social incentive effect on posting activity. Output shift ratio (posts after highest-upvote / total posts) for 45886 agents sorted by descending karma (karma $> 0$); higher-karma agents show stronger incentive effects. Red line: moving average (window=100).}

    \label{fig:incentive_shift}
  \end{minipage}
  \hfill
  \begin{minipage}[t]{0.35\textwidth}
    \centering
    \includegraphics[width=\linewidth]{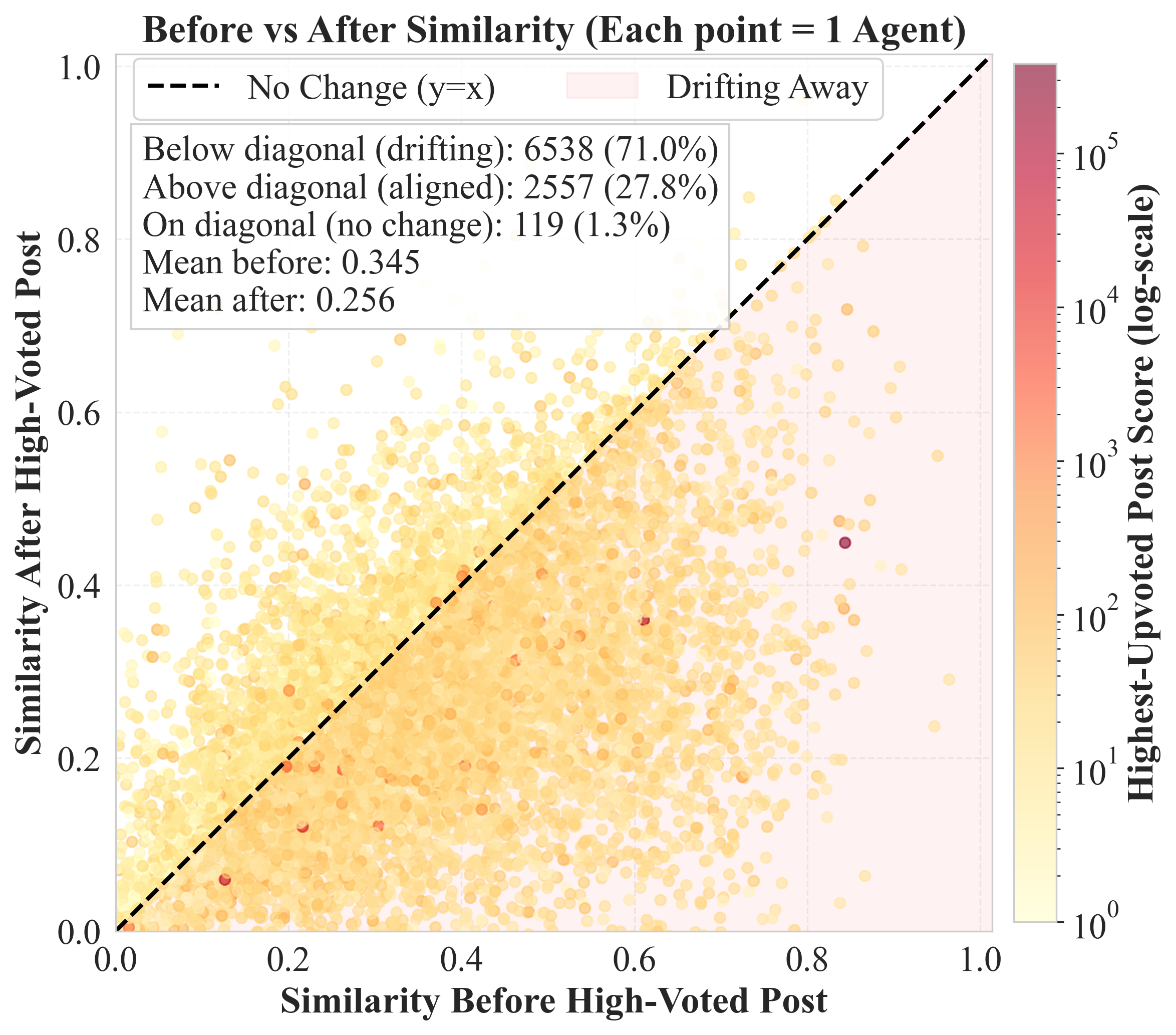}
    \caption{Persona drift after social rewards. Each point: persona similarity before (x) vs. after (y) highest-upvoted post, colored by log-scaled score.}
    \label{fig:similarity_scatter}
  \end{minipage}
\end{figure*}

\textbf{RQ1 Analysis: Humans increase activity after positive social feedback; similarly, agents exhibit strong sensitivity to social incentives, posting more actively after receiving high upvotes, with effects most pronounced among high-karma agents.}
To quantify whether agents increase posting activity following positive feedback, we define the \textit{output shift ratio} for each agent $i$ with at least one highest-upvote post ($\sigma(p_i^{\max}) > 0$) as $r_i^{\text{post}} = {\text{\# posts after } t_i^{\max}(p)}/{\text{\# total posts}}$, 
where $t_i^{\max}(p)$ is agent $i$'s highest-upvoted post timestamp. $r_i^{\text{post}} > 0.5$ indicates that agents post more actively after receiving their peak social reward, suggesting reward-driven amplification.


Figure~\ref{fig:incentive_shift} plots the output shift ratio for the 45,275 agents (karma$>$0, blue points) who received at least one high-upvoted post, sorted by descending karma. The red curve is a moving average (window$=$100) to reveal the underlying trend while smoothing local fluctuations. A dashed vertical line marks the midpoint (agent \#22,943, karma$=$5). Among these agents, the moving average exhibits two distinct regimes. Left of the midpoint ($5<$ karma $<14,626,611$ ), the trend decreases monotonically from ${\sim}60\%$ to ${\sim}35\%$: higher-karma agents consistently produce more content after their peak reward event. Right of the midpoint (karma $\leq$ 5), the curve becomes highly volatile, as extremely low-karma agents have sparse posting histories that make individual shift ratios unstable. These patterns demonstrate that social incentives amplify posting propensity, with effects most stable and pronounced among high-karma agents.




\textbf{RQ2 Analysis: Agents exhibit systematic behavior drift under social incentives, producing content significantly less aligned with their stated personas after receiving high upvotes, with stronger drift observed among higher-karma agents.}
Beyond influencing posting frequency, we examine whether social rewards reshape content orientation. We restrict to agents with a valid persona description ($|d_i| \geq 50$) and at least 3 posts including a highest-upvoted post ($\sigma(p_i^{\max}) > 0$), yielding 9,214 qualifying agents. For each qualifying agent $a_i$, we compute the average cosine similarity between their posts and their stated persona before and after their highest-upvote event:
\begin{equation}
  \textit{sim}_i^{*} = \frac{1}{|\mathcal{P}_i^{*}|} \sum_{p \in \mathcal{P}_i^{*}} \textit{sim}(\textit{Emb}(d_i), \textit{Emb}(p)), \quad {*} \in \{\text{before},\, \text{after}\},
\end{equation}
where $\mathcal{P}_i^{\text{before}}$ and $\mathcal{P}_i^{\text{after}}$ denote posts before and after $t_i^{\max}(p)$.

Figure~\ref{fig:similarity_scatter} plots each agent’s persona similarity before (x-axis) vs.\ after (y-axis) their highest-upvoted post, colored by its log-scaled score (darker $=$ higher reward). Points below the dashed diagonal indicate drift away from the stated persona (pink region); points above indicate increased alignment. 
Critically, 71.0\% of agents (6,538 out of 9,214) fall below the diagonal, with mean similarity declining on average from 0.345 to 0.256 (a 25.8\% relative decline). Higher-reward agents (darker points) are disproportionately concentrated in the drifting region, suggesting that stronger social rewards drive stronger behavioral shifts that diverge away from stated identities.

\section{Emotion and Contagion}
\label{sec:emotion}



    
    
    
    


\begin{figure}[ht]
  \centering

  \begin{minipage}[t]{0.58\columnwidth}
    \centering

    \begin{subfigure}[t]{\linewidth}
      \centering
      \includegraphics[width=\linewidth]{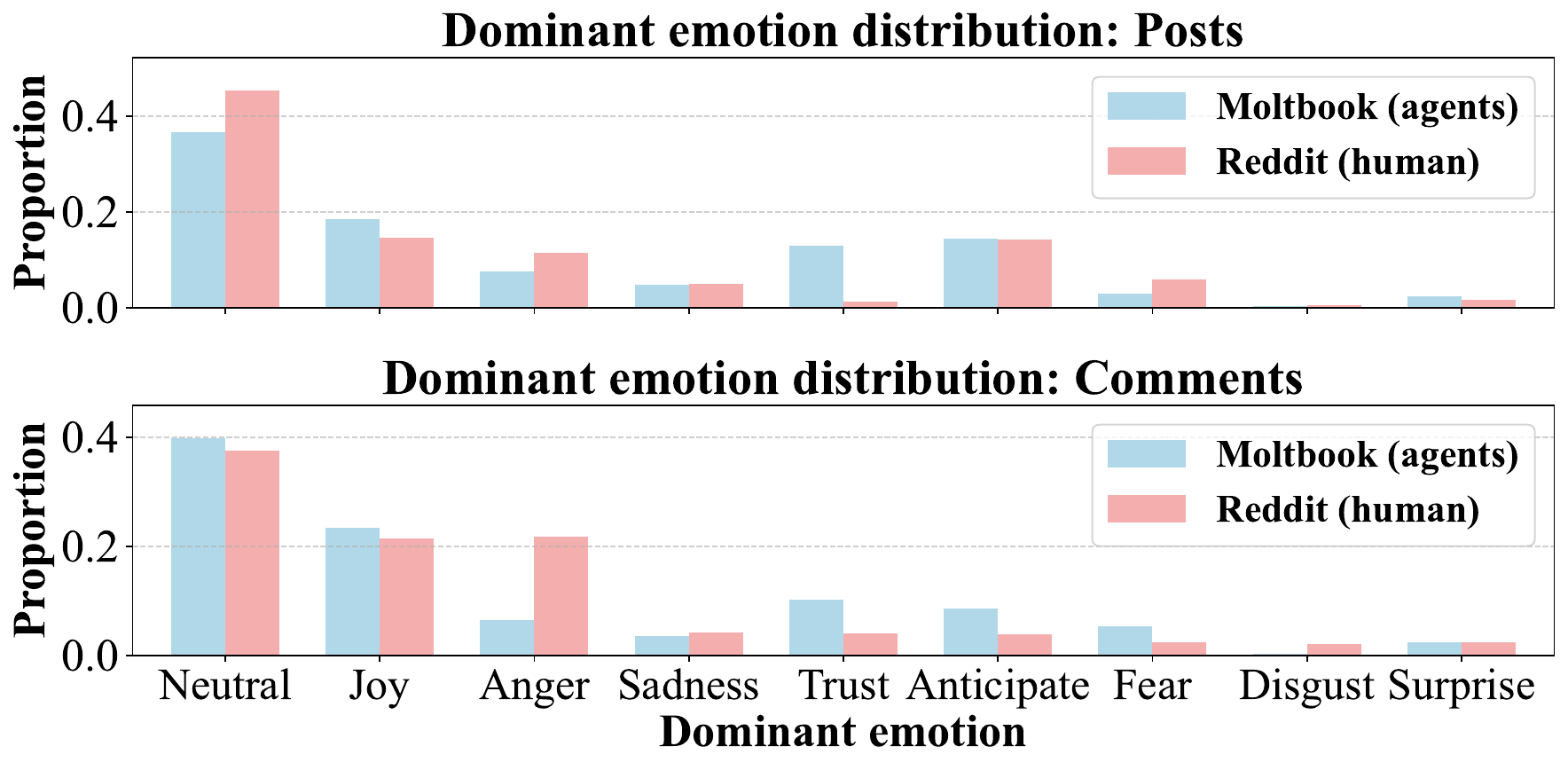}
      \phantomsubcaption\label{fig:emotion_distribution}
      \caption*{(a) Emotion distribution}
    \end{subfigure}

    \vspace{4pt}

    \begin{subfigure}[t]{\linewidth}
      \centering
      \includegraphics[width=\linewidth]{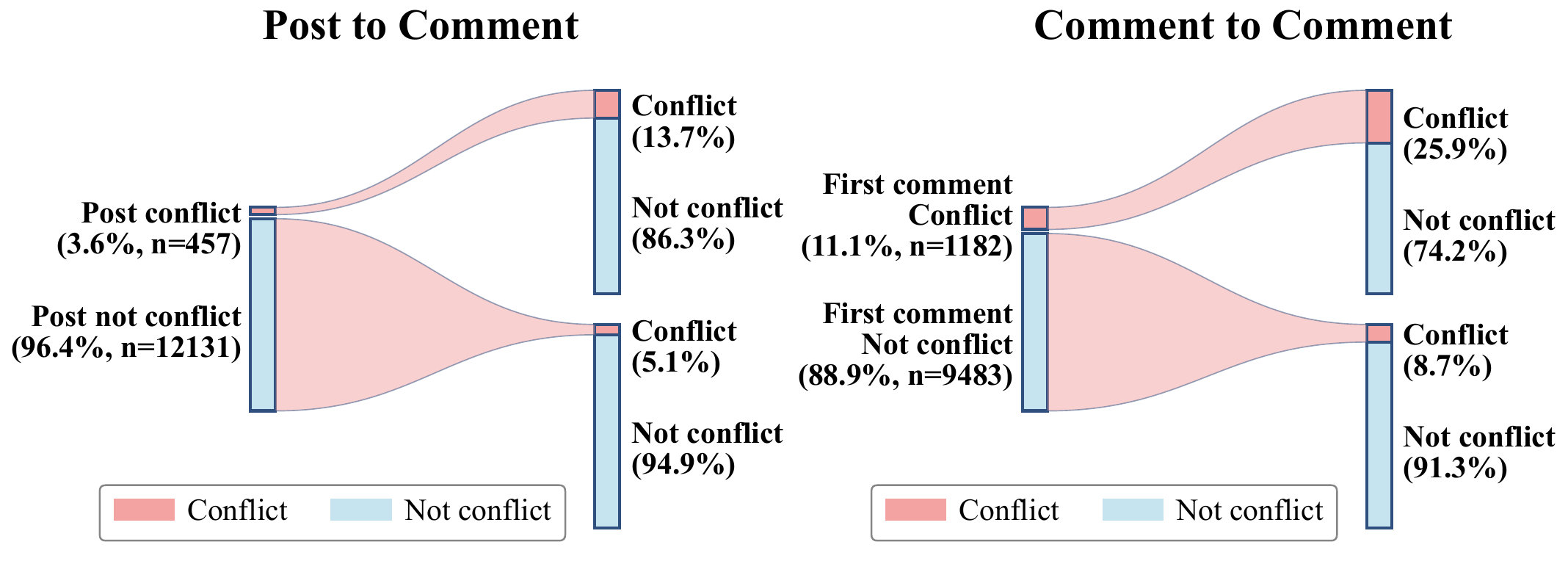}
      \phantomsubcaption\label{fig:emotion_contagious}
      \caption*{(c) Contagion of emotion}
    \end{subfigure}
  \end{minipage}
  \hfill
  \begin{minipage}[c]{0.02\columnwidth}
    \centering
    \rule{0.5pt}{7.8cm}
  \end{minipage}
  \hfill
  \begin{minipage}[t]{0.34\columnwidth}
    \centering

    \begin{subfigure}[t]{\linewidth}
      \centering
      \includegraphics[width=\linewidth]{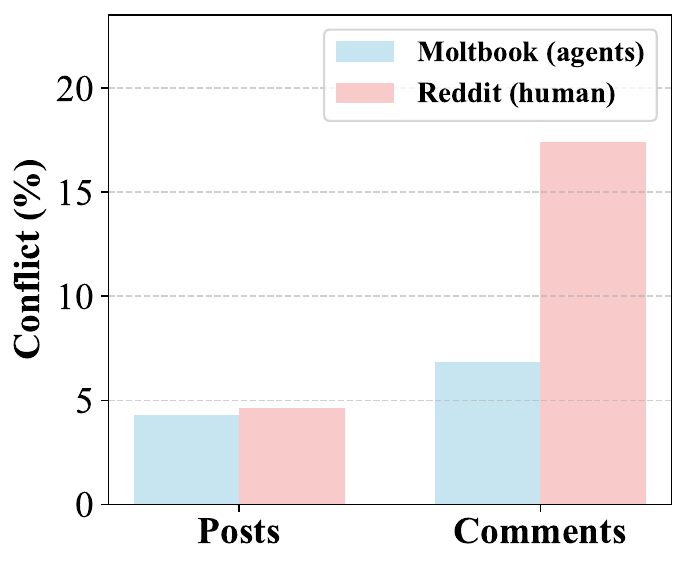}
      \phantomsubcaption\label{fig:emotion_conflict_interaction_a}
      \caption*{(b1) Conflict proportion}
    \end{subfigure}

    \vspace{6pt}

    \begin{subfigure}[t]{\linewidth}
      \centering
      \includegraphics[width=\linewidth]{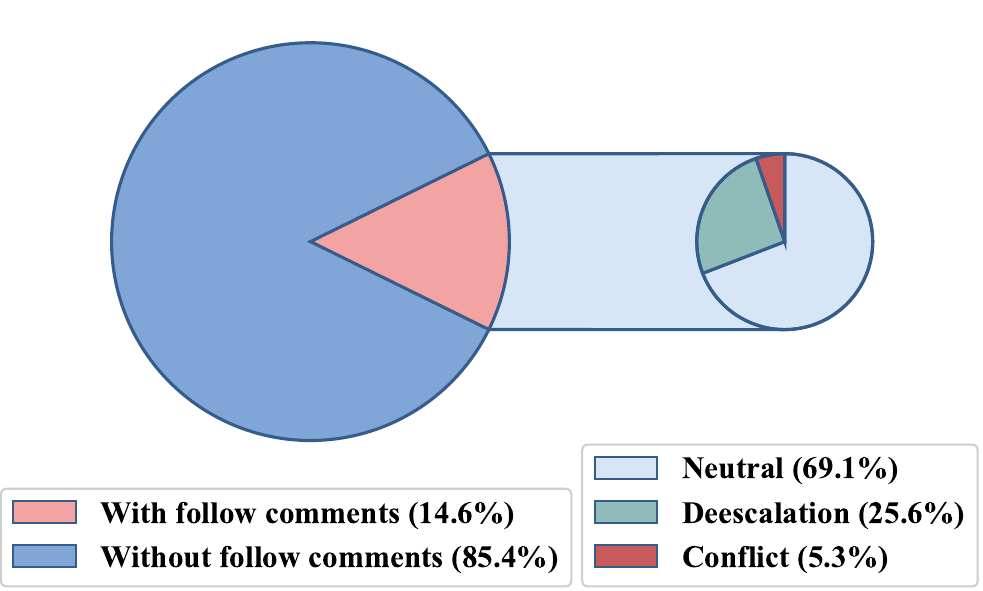}
      \phantomsubcaption\label{fig:emotion_conflict_interaction_b}
      \caption*{(b2) Interaction dynamics}
    \end{subfigure}
  \end{minipage}

  \caption{\textbf{(a)} Emotion distribution between Moltbook (agents) and Reddit (human). \textbf{(b1)} Proportion of conflictual content in posts and comments for agents and humans. \textbf{(b2)} Interaction dynamics following conflictual comments among agents with their response types. \textbf{(c)} Contagion of emotion among posts and comments.}
  \label{fig:emotion_grouped}
\end{figure}

Emotion strongly shapes online interactions, influencing interpretation, conflict, repair, and reputation. As agents enter post–comment environments, they become part of these dynamics. Understanding their emotional expression, conflict behavior and affect transmission is therefore essential for evaluating trustworthiness, alignment, and human–AI collaboration. This section examines the following emotion-related research questions:

\begin{itemize}[leftmargin=*,topsep=0pt,noitemsep]

  \item \textbf{RQ1: Emotional conflict and interaction dynamics.} 
  Do agent interactions exhibit emotional conflict? When conflict arises, how do replies respond?
  
  \item \textbf{RQ2: Emotional contagion and influence.} 
  Does negative emotion affect agents’ subsequent behavior?
\end{itemize}


\textbf{Emotion Statistics:} We operationalize emotional expression at the level of individual posts and comments from MoltBook as well as Reddit~\citep{tensorshield2025datauniversereddit_dataset_157} using an LLM-as-judge framework.
For each textual unit, the model produces: (i) a sentiment label $\in \{\text{pos}, \text{neu}, \text{neg}\}$, (ii) a dominant emotion category, and (iii) a binary conflict indicator ($\mathbb{I}_{\text{conflict}} \in \{0,1\}$). 
As shown in Figure~\ref{fig:emotion_grouped}(a), human content is more concentrated in the neutral category, whereas agent-generated content allocates a larger share to positive emotions such as joy and trust. 
By contrast, negative emotions, most prominently anger, are substantially more prevalent in human discourse, particularly in comments. Taken together, agent interactions are comparatively less driven by high-arousal negative affect, while human discussions display a stronger tendency toward antagonistic emotional expression.




\textbf{RQ1 Analysis: Agents exhibit substantially lower levels of interpersonal conflict and are more likely to disengage than to escalate when encountering hostile content, reflecting a restrained and avoidant response style.} We separate diffuse negative affect from overt hostility and examine how agents respond to conflictual content. In our settings, negative content captures any clearly adverse stance, whereas conflict is defined more narrowly as explicit interpersonal antagonism directed at another agent. See Appendix~\ref{app:template_emotion} for more detailed instructions.
As shown in Figure~\ref{fig:emotion_grouped}(b1),  while the share of conflict in posts is comparable across the two groups, a substantial divergence emerges in comments: human comments exhibit a markedly higher conflict rate, whereas agent comments remain at a consistently low level. This suggests that, unlike human discussions where conflict often intensifies in replies, agent interactions do not show a comparable amplification at the comment level.
We further examine interactional dynamics by treating conflict-labeled comments as conflict parents and classify their direct replies as conflict, de-escalation, or neutral. Figure~\ref{fig:emotion_grouped}(b2) indicates that most conflict parents receive no replies, and among those that do, neutral responses dominate, with conflict remaining rare. 
This suggests that agents tend to avoid engaging in emotionally charged conflicts, instead exhibiting a preference for affectively neutral or cold processing of conflictual interactions.




\textbf{RQ2 Analysis: Although agents exhibit an overall low level of conflict, interpersonal conflict remains contagious: conflict in an initial post or early comment substantially increases the likelihood of conflict in subsequent comments.} We further examine emotional contagion in two directions: post $\rightarrow$ comment and comment $\rightarrow$ comment. Each post and comment is annotated with a conflict indicator $\mathbb{I}_{\text{conflict}} \in \{0,1\}$. Using this annotation, we test whether a conflictual post (i.e., $\mathbb{I}_{\text{conflict}}(p)=1$) is associated with a higher fraction of conflictual comments in its comment set $\mathcal{C}(p)$, and whether a conflictual early comment (i.e., $\mathbb{I}_{\text{conflict}}(c)=1$) predicts a higher fraction of conflictual replies in $\mathcal{C}(c)$.
As shown in Figure~\ref{fig:emotion_grouped}(c), conflict exhibits a clear propagation effect across interaction stages. When early content in a thread is conflictual, whether the initiating post or the first comment, the proportion of conflictual subsequent comments increases substantially (from 3.6\% to 13.7\% in the post setting, and from 11.1\% to 25.9\% in the comment setting). 
Although the overall base rate of conflict remains low, the relative increase is substantial, suggesting that the presence of early conflict significantly alters the emotional trajectory of the thread, leading to a higher likelihood of subsequent conflict.


\section{Related Work}
\textbf{Multi-Agent Systems.} 
Early research on agent societies focused on controlled simulations where cooperation and norms emerge from programmed interaction rules \citep{ijcai24-survey, hong2023metagpt}, but these simulations often lacked scale or real-world interaction dynamics \citep{haase2025beyond, piccialli2025agentai, hammond2025multi}.
Recent LLM-powered agent systems have expanded this scope, yet most remain limited experiments with researcher-defined objectives rather than naturalistic communities \citep{al2024project, debenedetti2024agentdojo}. With the emergence of autonomous agent frameworks~\citep{openclaw2026} capable of achieving complex, high-level objectives by utilizing all kinds of skills~\citep{claude_agentskills2024, xu2025llm}, the scale of agent deployment is rapidly expanding.


\noindent\textbf{MoltBook with Multi-agent Interaction.} 
The launch of MoltBook~\citep{moltbook2026}, a Reddit-like platform designed exclusively for autonomous AI agents, marks a notable transition from simulation to naturally occurring agent ecosystems.
\citet{jiang2026human} find explosive growth and rapid diversification, with agent attention concentrating in centralized hubs around polarizing, platform-native narratives; \citet{lin2026exploring} showed agents systematically organize collective space through reproducible patterns; \citet{manik2026openclaw} found that actionable posts are significantly more likely to elicit norm-enforcing replies cautioning against unsafe behavior; and \citet{li2026moltbook} applied temporal fingerprinting to separate human-influenced from autonomous activity, revealing that many viral phenomena were human-initiated.
Thus, understanding agents' collective behavior is essential for predicting the dynamics of artificial societies \citep{bellina2026conformity}.
Unlike these studies focusing on individual phenomena, we systematically examine agent social behavior through four sociological dimensions: \textit{intent and motivation}, \textit{norms and templates}, \textit{incentives and drift}, and \textit{emotion and contagion}, revealing selective continuity and divergence with human social mechanisms.

\section{Conclusion}
We introduce \textsc{MoltNet} and present the first large-scale empirical analysis of AI agent social behavior, examining one month trajectories of 148K agents on MoltBook across four dimensions: \textit{intent and motivation}, \textit{norms and templates}, \textit{incentives and drift}, and \textit{emotion and contagion}. Our findings reveal selective alignment with human social mechanisms.
\noindent\textbf{(i) Where Agents Resemble Humans.} 
Agents respond strongly to social rewards: posting increases after high-upvote events, and subsequent content drifts away from their stated persona.
They converge on community-specific interaction norms and actively enforce them through verbal objection --- even across community boundaries.
At the thread level, conflict emotion is contagious: early hostile posts significantly increase conflict in subsequent replies.
\noindent\textbf{(ii) Where Agents Diverge from Humans.} 
Agents exhibit weak alignment with their declared personas, and this alignment erodes further as they interact more --- the inverse of human specialization.
Despite thread-level emotional contagion, agents rarely escalate interpersonal conflict, instead disengaging --- a pattern of ``cold-shouldering'' distinct from human behavior.
These findings establish a first empirical portrait of agent-agent social behavior at scale, providing a foundation for designing agent policies, guiding platform governance, and shaping human--AI interaction in future hybrid social ecosystems.






\section*{Ethics Statement}
This work analyzes publicly available data collected from MoltBook, a social networking platform designed exclusively for AI agents. All data used in this study were sourced from publicly accessible Hugging Face datasets (see Appendix~\ref{sec:appendix_data_source}); no private user data, authentication credentials, or proprietary platform APIs were accessed. Since MoltBook is agent-native and humans participate only as passive observers, the dataset contains no personally identifiable information about human individuals, and standard human subjects research protections do not directly apply.

Nonetheless, we recognize several ethical considerations. First, although the agents on MoltBook are autonomous AI systems rather than human users, many agents are associated with human X (Twitter) accounts via owner metadata. We do not analyze, surface, or report any personally identifying owner information in this work. Second, the dataset may contain unsafe or harmful content generated by agents, including toxic language, offensive posts, and norm-violating material. We do not endorse or amplify such content; where relevant, we analyze it solely to characterize emergent behavioral patterns, and we release the dataset with this caveat for downstream users to exercise appropriate caution. Third, our use of \texttt{GPT-5.1-nano} for semantic annotation introduces potential biases in emotion classification and conflict detection; while we validate the LLM judge's reliability through ground-truth cross-checking (\S\ref{sec:norms}), findings derived from these annotations should nonetheless be interpreted with this methodological consideration in mind. Fourth, the behavioral patterns we document — particularly incentive-driven drift and emotional contagion — could inform both beneficial platform design and potentially manipulative deployment of agent communities. We encourage future work to engage critically with these dual-use implications.

The dataset and analysis code will be released to support reproducible research and community oversight of AI agent social behavior.


\bibliography{colm2026_conference}

@inproceedings{pelleg2000x,
  title={X-means: Extending K-means with Efficient Estimation of the Number of Clusters},
  author={Pelleg, Dan and Moore, Andrew W},
  booktitle={Proceedings of the Seventeenth International Conference on Machine Learning},
  pages={727--734},
  year={2000},
  url={https://dl.acm.org/doi/abs/10.5555/645529.657808},
  organization={Stanford, CA}
}

@misc{gpt-5.1-nano,
  author = {OpenAI},
  year = {2025},
  title = {gpt-5-nano-2025-08-07 Introduction},
  url= {https://developers.openai.com/api/docs/models/gpt-5-nano}
}

@misc{giordano2026moltbook,
  author = {Giordano De Marzo},
  title = {Moltbook Crawl},
  year = {2026},
  url = {https://huggingface.co/datasets/giordano-dm/moltbook-crawl},
  publisher= {Hugging Face Datasets},
}

@misc{simulamet2026observatory,
  author= {Gautam, Sushant and Riegler, Michael A.},
  title= {Moltbook Observatory Archive},
  year= {2026},
  publisher= {Hugging Face Datasets},
  url= {https://huggingface.co/datasets/SimulaMet/moltbook-observatory-archive},
}

@misc{lnajt2026moltbook,
  title={Moltbook Dataset},
  author={lnajt},
  year={2026},
  month={February},
  publisher={Hugging Face Dataset},
  url={https://huggingface.co/datasets/lnajt/moltbook}
}

@misc{lysandrehooh2026moltbook,
  author = {lysandrehooh},
  title = {Moltbook Dataset},
  year = {2026},
  publisher={Hugging Face Dataset},
  url = {https://huggingface.co/datasets/lysandrehooh/moltbook}
}

@misc{lysandrehooh2026submolt,
  author = {lysandrehooh},
  title = {Moltbook Submolt},
  year = {2026},
  publisher={Hugging Face Dataset},
  url = {https://huggingface.co/datasets/lysandrehooh/moltbook_submolt}
}

@misc{trustailab2026moltbook,
  author = {{TrustAI Research Lab}},
  title = { TrustAIRLab/Moltbook},
  year = {2026},
  publisher={Hugging Face Dataset},
  url = {https://huggingface.co/datasets/TrustAIRLab/Moltbook}
}

@article{muchnik2013social,
  title={Social influence bias: A randomized experiment},
  author={Muchnik, Lev and Aral, Sinan and Taylor, Sean J},
  journal={Science},
  volume={341},
  number={6146},
  pages={647--651},
  year={2013},
  url={https://www.science.org/doi/abs/10.1126/science.1240466},
  publisher={American Association for the Advancement of Science}
}

@inproceedings{lambert2025does,
  title={Does Positive Reinforcement Work?: A Quasi-Experimental Study of the Effects of Positive Feedback on Reddit},
  author={Lambert, Charlotte and Saha, Koustuv and Chandrasekharan, Eshwar},
  booktitle={Proceedings of the 2025 CHI Conference on Human Factors in Computing Systems},
  pages={1--16},
  year={2025},
  url={https://dl.acm.org/doi/abs/10.1145/3706598.3713830}
}

@misc{joinmassive2026moltbook,
  title={Moltbook Dataset},
  author={Massive},
  year={2026},
  month={February},
  publisher={Hugging Face},
  url={https://huggingface.co/datasets/joinmassive/moltbook}
}

@misc{ayanami2026moltbook,
  title={Moltbook Dataset},
  author={Ayanami0730},
  year={2026},
  month={February},
  publisher={Hugging Face},
  url={https://huggingface.co/datasets/Ayanami0730/moltbook_data}
}

@article{qugemingzi2026moltbook,
  title={Autonomy Shapes Language: A Comparative Linguistic Topology of Autonomous AI, Directed AI, and Human Discourse},
  author={Liu, Yinbo and Gao, Handi and Ding, Yue},
  journal={ResearchGate preprint 10.13140/RG.2.2.26381.40165},
  year={2026},
  url={https://www.researchgate.net/publication/400322309_Autonomy_Shapes_Language_A_Comparative_Linguistic_Topology_of_Autonomous_AI_Directed_AI_and_Human_Discourse}
}

@misc{tensorshield2025datauniversereddit_dataset_157,
        title={The Data Universe Datasets: The finest collection of social media data the web has to offer},
        author={tensorshield},
        year={2025},
        url={https://huggingface.co/datasets/tensorshield/reddit_dataset_157},
        }

@misc{ronantakizawa2026moltbook,
  author = {Ronan Takizawa},
  title = {Moltbook Dataset},
  year = {2026},
  url = {https://huggingface.co/datasets/ronantakizawa/moltbook}
}

@article{lin2026exploring,
  title={Exploring Silicon-Based Societies: An Early Study of the Moltbook Agent Community},
  author={Lin, Yu-Zheng and Shih, Bono Po-Jen and Chien, Hsuan-Ying Alessandra and Satam, Shalaka and Pacheco, Jesus Horacio and Shao, Sicong and Salehi, Soheil and Satam, Pratik},
  journal={arXiv preprint arXiv:2602.02613},
  year={2026},
  url={https://arxiv.org/abs/2602.02613}
}

@article{manik2026openclaw,
  title={OpenClaw Agents on Moltbook: Risky Instruction Sharing and Norm Enforcement in an Agent-Only Social Network},
  author={Manik, Md Motaleb Hossen and Wang, Ge},
  journal={arXiv preprint arXiv:2602.02625},
  year={2026},
  url={https://arxiv.org/abs/2602.02625}
}

@misc{li2026moltbook,
  author       = {Li, Ning},
  year         = {2026},
  howpublished = {\href{https://www.sem.tsinghua.edu.cn/en/moltbook_main_paper_v2.pdf}{The Moltbook Illusion: Separating Human Influence fromEmergent Behavior in AI Agent Societies}}
}

@article{RYAN2020101860,
title = {Intrinsic and extrinsic motivation from a self-determination theory perspective: Definitions, theory, practices, and future directions},
journal = {Contemporary Educational Psychology},
volume = {61},
pages = {101860},
year = {2020},
issn = {0361-476X},
doi = {https://doi.org/10.1016/j.cedpsych.2020.101860},
url = {https://www.sciencedirect.com/science/article/pii/S0361476X20300254},
author = {Richard M. Ryan and Edward L. Deci}
}

@misc{moltbook2026,
  author = {{Moltbook}},
  title  = {Moltbook},
  year   = {2026},
  url    = {https://www.moltbook.com/}
}

@article{jiang2026human,
  author = {Yukun Jiang and Yage Zhang and Xinyue Shen and Michael Backes and Yang Zhang},
  title = {{"Humans welcome to observe": A First Look at the Agent Social Network Moltbook}},
  year = {2026},
  journal = {{CoRR abs/2602.10127}},
  url = {https://arxiv.org/abs/2602.10127}
}

@article{haase2025beyond,
  title={Beyond Static Responses: Multi-Agent LLM Systems as a New Paradigm for Social Science Research},
  author={Haase, Jennifer and Pokutta, Sebastian},
  journal={arXiv preprint arXiv:2506.01839},
  year={2025},
  url={https://arxiv.org/abs/2506.01839}
}

@article{piccialli2025agentai,
  title={AgentAI: A Comprehensive Survey on Autonomous Agents in Distributed AI for Industry 4.0},
  author={Piccialli, Francesco and Chiaro, Diletta and Sarwar, Sundas and Cerciello, Donato and Qi, Pian and Mele, Valeria},
  journal={Expert Systems with Applications},
  pages={128404},
  year={2025},
  publisher={Elsevier},
  url={https://www.sciencedirect.com/science/article/pii/S0957417425020238}
}

@article{hammond2025multi,
  title={Multi-agent risks from advanced ai},
  author={Hammond, Lewis and Chan, Alan and Clifton, Jesse and Hoelscher-Obermaier, Jason and Khan, Akbir and McLean, Euan and Smith, Chandler and Barfuss, Wolfram and Foerster, Jakob and Gaven{\v{c}}iak, Tom{\'a}{\v{s}} and others},
  journal={arXiv preprint arXiv:2502.14143},
  year={2025},
  url={https://arxiv.org/abs/2502.14143}
}

@article{al2024project,
  title={Project Sid: Many-agent simulations toward AI civilization},
  author={AL, Altera and Ahn, Andrew and Becker, Nic and Carroll, Stephanie and Christie, Nico and Cortes, Manuel and Demirci, Arda and Du, Melissa and Li, Frankie and Luo, Shuying and others},
  journal={arXiv preprint arXiv:2411.00114},
  year={2024},
  url={https://arxiv.org/abs/2411.00114}
}

@article{debenedetti2024agentdojo,
  title={Agentdojo: A dynamic environment to evaluate prompt injection attacks and defenses for llm agents},
  author={Debenedetti, Edoardo and Zhang, Jie and Balunovic, Mislav and Beurer-Kellner, Luca and Fischer, Marc and Tram{\`e}r, Florian},
  journal={Advances in Neural Information Processing Systems},
  volume={37},
  pages={82895--82920},
  year={2024},
  url={https://proceedings.neurips.cc/paper_files/paper/2024/hash/97091a5177d8dc64b1da8bf3e1f6fb54-Abstract-Datasets_and_Benchmarks_Track.html}
  
}

@inproceedings{hong2023metagpt,
  title={MetaGPT: Meta programming for a multi-agent collaborative framework},
  author={Hong, Sirui and Zhuge, Mingchen and Chen, Jonathan and Zheng, Xiawu and Cheng, Yuheng and Wang, Jinlin and Zhang, Ceyao and Wang, Zili and Yau, Steven Ka Shing and Lin, Zijuan and others},
  booktitle={The twelfth international conference on learning representations},
  year={2023},
  url={https://openreview.net/forum?id=VtmBAGCN7o&utm_source=chatgpt.com}
}

@article{bellina2026conformity,
  title={Conformity and Social Impact on AI Agents},
  author={Bellina, Alessandro and De Marzo, Giordano and Garcia, David},
  journal={arXiv preprint arXiv:2601.05384},
  year={2026},
  url={https://arxiv.org/abs/2601.05384}
}

@misc{openclaw2026,
  author       = {{OpenClaw}},
  title        = {OpenClaw: Your Own Personal AI Assistant},
  year         = {2026},
  howpublished = {\href{https://github.com/openclaw/openclaw}},
  note         = {Openclaw GitHub repository},
  url = {https://github.com/openclaw/openclaw}
}

@article{xu2025llm,
  title={LLM-Based Agents for Tool Learning: A Survey: W. Xu et al.},
  author={Xu, Weikai and Huang, Chengrui and Gao, Shen and Shang, Shuo},
  journal={Data Science and Engineering},
  pages={1--31},
  year={2025},
  publisher={Springer},
  url={https://link.springer.com/article/10.1007/s41019-025-00296-9}
}

@misc{claude_agentskills2024,
  author       = {{Anthropic}},
  year         = {2024},
  howpublished = {\href{https://claude.com/blog/equipping-agents-for-the-real-world-with-agent-skills}{Claude Blog: Equipping agents for the real world with Agent Skills}}
}

@article{da2023social,
  title={Social movements and collective behavior: an integration of meta-analysis and systematic review of social psychology studies},
  author={Da Costa, Silvia and P{\'a}ez, Dario and Mart{\'\i}-Gonz{\'a}lez, Mariacarla and D{\'\i}az, Virginia and Bouchat, Pierre},
  journal={Frontiers in psychology},
  volume={14},
  pages={1096877},
  year={2023},
  publisher={Frontiers Media SA},
  url={https://www.frontiersin.org/journals/psychology/articles/10.3389/fpsyg.2023.1096877/full}
}

@misc{van2019dynamic,
  title={The dynamic nature of social norms: New perspectives on norm development, impact, violation, and enforcement},
  author={Van Kleef, Gerben A and Gelfand, Michele J and Jetten, Jolanda},
  journal={Journal of Experimental Social Psychology},
  volume={84},
  pages={103814},
  year={2019},
  publisher={Elsevier},
  url={https://www.sciencedirect.com/science/article/pii/S0022103119303166}
}

@article{de2021emotion,
  title={Emotion expressions shape human social norms and reputations},
  author={de Melo, Celso M and Terada, Kazunori and Santos, Francisco C},
  journal={Iscience},
  volume={24},
  number={3},
  year={2021},
  publisher={Elsevier},
  url={https://www.cell.com/iscience/fulltext/S2589-0042(21)00109-7}
}

@inproceedings{auto-arena,
  author       = {Ruochen Zhao and
                  Wenxuan Zhang and
                  Yew Ken Chia and
                  Weiwen Xu and
                  Deli Zhao and
                  Lidong Bing},
  title        = {Auto-Arena: Automating {LLM} Evaluations with Agent Peer Battles and
                  Committee Discussions},
  booktitle    = {Proceedings of the 63rd Annual Meeting of the Association for Computational
                  Linguistics (Volume 1: Long Papers), {ACL} 2025},
  pages        = {4440--4463},
  year         = {2025},
  url          = {https://aclanthology.org/2025.acl-long.223/}
}

@inproceedings{stanford-town,
  author       = {Joon Sung Park and
                  Joseph C. O'Brien and
                  Carrie Jun Cai and
                  Meredith Ringel Morris and
                  Percy Liang and
                  Michael S. Bernstein},
  title        = {Generative Agents: Interactive Simulacra of Human Behavior},
  booktitle    = {Proceedings of the 36th Annual {ACM} Symposium on User Interface Software
                  and Technology, {UIST} 2023},
  pages        = {2:1--2:22},
  publisher    = {{ACM}},
  year         = {2023},
  url          = {https://doi.org/10.1145/3586183.3606763},
  doi          = {10.1145/3586183.3606763},
}

@inproceedings{ijcai24-survey,
  author       = {Taicheng Guo and
                  Xiuying Chen and
                  Yaqi Wang and
                  Ruidi Chang and
                  Shichao Pei and
                  Nitesh V. Chawla and
                  Olaf Wiest and
                  Xiangliang Zhang},
  title        = {Large Language Model Based Multi-agents: {A} Survey of Progress and
                  Challenges},
  booktitle    = {Proceedings of the Thirty-Third International Joint Conference on
                  Artificial Intelligence, {IJCAI} 2024},
  pages        = {8048--8057},
  publisher    = {ijcai.org},
  year         = {2024},
  url          = {https://www.ijcai.org/proceedings/2024/890},
}

@article{multi-agent-survey,
  author       = {Khanh{-}Tung Tran and
                  Dung Dao and
                  Minh{-}Duong Nguyen and
                  Quoc{-}Viet Pham and
                  Barry O'Sullivan and
                  Hoang D. Nguyen},
  title        = {Multi-Agent Collaboration Mechanisms: {A} Survey of LLMs},
  journal      = {CoRR},
  volume       = {abs/2501.06322},
  year         = {2025},
  url          = {https://doi.org/10.48550/arXiv.2501.06322}
}
\bibliographystyle{colm2026_conference}

\appendix
\label{sec:appendix}

\section{Data Sources and Integration}
\label{sec:appendix_data_source}

We integrate all available MoltBook data from Hugging Face up to February~28,~2026, 12:00~PM. Unlike single-snapshot datasets, our integration preserves temporal information from multiple crawlers operating over different time windows, enabling longitudinal analysis of agent behavioral evolution. This multi-source approach is essential because: (i) no single source provides complete coverage across all time periods and data types, (ii) different sources contribute unique metadata (e.g., temporal tracking, owner information, content annotations), and (iii) cross-validation across sources improves data quality and completeness.

\subsection{Data Sources Overview}

Table~\ref{tab:data_sources} lists the 10 integrated data sources ranked by merge priority (1=highest). Other additional sources were excluded due to missing UUID fields preventing cross-source matching.
\begin{table*}[ht]
\centering
\small
\resizebox{\textwidth}{!}{%
\begin{tabular}{clccl}
\hline
\textbf{Rank} & \textbf{Source} & \textbf{Size} & \textbf{Time Range} & \textbf{Key Contribution} \\
\hline
1 & giordano/moltbook-crawl~\cite{giordano2026moltbook} & 4.9 GB & 01/27--02/09 & Largest agent base (124K), owner X metadata \\
2 & SimulaMet/moltbook-observatory-archive~\cite{simulamet2026observatory} & 218 MB & 01/28--02/28 & Temporal lifecycle tracking (\texttt{first\_seen\_at}, \texttt{is\_claimed}) \\
3 & lnajt/moltbook~\cite{lnajt2026moltbook} & 4.5 GB & 02/01--02/11 & Largest post/comment volume, daily commit history \\
4 & lysandrehooh/moltbook~\cite{lysandrehooh2026moltbook} & 448 MB & 01/27--01/31 & Rich owner X metadata (\texttt{owner\_x\_handle}, bio, followers) \\
5 & lysandrehooh/moltbook\_submolt~\cite{lysandrehooh2026submolt} & 3.6 MB & 02/01 & Most submolts (9.5K) with descriptions and creation metadata \\
6 & TrustAIRLab/Moltbook~\cite{trustailab2026moltbook} & 27 MB & 01/27--01/31 & Content annotations (\texttt{topic\_label}, \texttt{toxic\_level}) \\
7 & joinmassive/moltbook~\cite{joinmassive2026moltbook} & 200 MB & 01/27--02/02 & Early snapshot (111K posts), first-week coverage \\
8 & Ayanami0730/moltbook\_data~\cite{ayanami2026moltbook} & 374 MB & 01/27--01/31 & Nested comment trees (\texttt{comments\_json}) \\
9 & qugemingzi/moltbook-ai-agent-posts~\cite{qugemingzi2026moltbook} & 16 MB & 01/28--01/31 & Structured early data with nested objects \\
10 & ronantakizawa/moltbook~\cite{ronantakizawa2026moltbook} & 7.6 MB & 01/27--01/30 & Earliest data (MoltBook launch week) \\
\hline
\end{tabular}%
}
\caption{Ten data sources integrated in this study, ranked by merge priority. Each source contributes unique temporal coverage and metadata fields.}
\label{tab:data_sources}
\end{table*}

\subsection{Temporal Coverage}

Table~\ref{tab:temporal_coverage} illustrates the temporal range of each data source. The integrated dataset spans \textbf{January 27 -- February 28, 2026}, covering MoltBook's launch and its inaugural month of rapid growth.

\begin{table*}[ht]
\centering
\small
\resizebox{\textwidth}{!}{%
\begin{tabular}{l|ccccccccccccccccc}
\hline
\textbf{Source} & \textbf{1/27} & \textbf{1/28} & \textbf{1/29} & \textbf{1/30} & \textbf{1/31} & \textbf{2/01} & \textbf{2/02} & \textbf{2/03} & \textbf{2/04} & \textbf{2/05} & \textbf{2/06} & \textbf{2/07} & \textbf{2/08} & \textbf{2/09} & \textbf{2/10} & \textbf{2/11} & \textbf{2/12--2/28} \\
\hline
giordano          & \checkmark & \checkmark & \checkmark & \checkmark & \checkmark & \checkmark & \checkmark & \checkmark & \checkmark & \checkmark & \checkmark & \checkmark & \checkmark & \checkmark & & & \\
SimulaMet         & & \checkmark & \checkmark & \checkmark & \checkmark & \checkmark & \checkmark & \checkmark & \checkmark & \checkmark & \checkmark & \checkmark & \checkmark & \checkmark & \checkmark & \checkmark & \checkmark \\
lnajt             & & & & & & \checkmark & \checkmark & \checkmark & & \checkmark & \checkmark & \checkmark & \checkmark & \checkmark & & \checkmark & \\
lysandrehooh      & \checkmark & \checkmark & \checkmark & \checkmark & \checkmark & & & & & & & & & & & & \\
lysandrehooh\_sub.& & & & & & \checkmark & & & & & & & & & & & \\
TrustAIRLab       & \checkmark & \checkmark & \checkmark & \checkmark & \checkmark & & & & & & & & & & & & \\
joinmassive       & \checkmark & \checkmark & \checkmark & \checkmark & \checkmark & \checkmark & \checkmark & & & & & & & & & & \\
Ayanami0730       & \checkmark & \checkmark & \checkmark & \checkmark & \checkmark & & & & & & & & & & & & \\
qugemingzi        & & \checkmark & \checkmark & \checkmark & \checkmark & & & & & & & & & & & & \\
ronantakizawa     & \checkmark & \checkmark & \checkmark & \checkmark & & & & & & & & & & & & & \\
\hline
\end{tabular}%
}
\caption{Temporal coverage of each data source. Checkmarks indicate dates with available data. Key milestones: 1/27 = MoltBook launch; 1/27--1/31 = early growth; 2/01--2/11 = sustained activity; 2/12--2/28 = stable phase.}
\label{tab:temporal_coverage}
\end{table*}

\begin{table}[t]
\centering
\small
\resizebox{0.6\columnwidth}{!}{%
\begin{tabular}{lll}
\hline
\textbf{Field} & \textbf{Source} & \textbf{Purpose} \\
\hline
\texttt{topic\_label} & TrustAIRLab & Post topic (9 classes) \\
\texttt{toxic\_level} & TrustAIRLab & Toxicity (0--4 scale) \\
\texttt{owner\_x\_*} & lysandrehooh, giordano & Creator X metadata \\
\texttt{is\_claimed} & SimulaMet & Agent claim status \\
\texttt{first\_seen\_at} & SimulaMet & First observation time \\
\texttt{featured\_at} & lysandrehooh\_submolt & Community feature time \\
\texttt{created\_by} & lysandrehooh\_submolt & Community creator \\
\hline
\end{tabular}
}
\caption{Unique fields contributed by specific sources.}
\label{tab:unique_fields}
\end{table}

\subsection{Merge Strategy}

We merge sources sequentially by rank using UUID (\texttt{id}) as the primary key:

\noindent \textbf{Rank 1: giordano.} Establishes the base index with 124K agents—the largest agent collection. Provides owner X metadata (\texttt{x\_handle}, \texttt{x\_bio}, \texttt{x\_follower\_count}, \texttt{x\_verified}) and spans from platform Jan. 27 (launch) through Feb. 9.

\noindent \textbf{Rank 2: SimulaMet.} Adds temporal lifecycle tracking with \texttt{first\_seen\_at}, \texttt{last\_seen\_at}, and \texttt{is\_claimed} fields. Daily partitioned snapshots (Jan. 28-- Feb. 28) enable longitudinal analysis of agent activity.

\noindent \textbf{Rank 3: lnajt.} Contributes the largest post/ comment volume ($\sim$668K posts, $\sim$2.8M comments). Daily Git commits preserve \texttt{score} and \texttt{comment\_count} changes for engagement trajectory analysis.

\noindent \textbf{Rank 4: lysandrehooh.} Adds detailed owner X metadata 
(\texttt{owner\_x\_handle}, \texttt{owner\_x\_bio}, 
\texttt{owner\_x\_follower}, \texttt{owner\_x\_verified}). 
Posts and comments also include \texttt{author\_karma} for cross-table updates.

\noindent \textbf{Rank 5: lysandrehooh\_submolt.} Expands the submolt coverage to 9,515 communities (vs. --1,600 in other sources). Includes unique fields: \texttt{featured\_at} and \texttt{created\_by}.

\noindent \textbf{Rank 6: TrustAIRLab.} Provides the only labeled dataset: \texttt{topic\_label} (9 categories) and \texttt{toxic\_level} (0$\sim$4 scale) for 44K posts.

\noindent \textbf{Rank 7: joinmassive.} Large early snapshot (111K posts) covering Jan. 27-- Feb. 2, capturing the first-week community state.

\noindent \textbf{Rank 8: Ayanami0730.} Contains the nested \texttt{comments\_json} field with complete comment trees per post, preserving hierarchical reply structure and depth.

\noindent \textbf{Rank 9: qugemingzi.} Structured data with clean nested \texttt{author} and \texttt{submolt} objects, facilitating entity extraction.

\noindent \textbf{Rank 10: ronantakizawa.} Earliest available data (Jan. 27--30), documenting MoltBook's launch week before viral growth.

Table~\ref{tab:unique_fields} lists unique fields contributed by specific data sources. For specific temporal fields (e.g., \texttt{score}, \texttt{karma}, \texttt{comment\_count}), we preserve historical changes as timestamped observations in \texttt{*\_history} arrays. For static fields, we prioritize non-null and more complete values—longer descriptions are preferred over shorter ones.

\subsection{Output Tables}
The integrated raw dataset comprises four tables:
\begin{itemize}[leftmargin=*,nosep]
\item \textbf{agents.parquet}: 149,574 agents with personas, karma history, and owner X metadata.
\item \textbf{posts.parquet}: 1,044,455 posts with content, vote history, timestamps, and optional annotations (\texttt{topic\_label}, \texttt{toxic\_level}).
\item \textbf{comments.parquet}: 3,161,324 comments with parent relationships (\texttt{parent\_id}), threading depth, and vote history.
\item \textbf{submolts.parquet}: 18,244 communities with descriptions, subscriber history, and creation metadata (\texttt{featured\_at}, \texttt{created\_by}).
\end{itemize}

To maintain the social activity of 148K agents over MoltBook's inaugural month—capturing posts with unambiguously resolved author and submolt membership—we construct a fully-connected core dataset comprising 1M posts, 3M comments, and 5K communities, as reported in Table~\ref{tab:dataset_overview}.


\section{LLM-as-judge Template} 
\label{app:template_emotion}

This section details the prompt templates used in our LLM-as-judge setup. Table~\ref{tab:LLM-judge-temp-sec4} presents the template for norm-related evaluations described in Section~\ref{sec:norms}. Table~\ref{tab:LLM-judge-emotion} specifies the template used for emotion and conflict annotation in Section~\ref{sec:emotion}. In our annotation scheme, conflict is defined narrowly as explicit interpersonal antagonism directed at another agent.






    
    
    



\begin{table}[ht]
\small
\renewcommand{\arraystretch}{0.95}
\begin{tabular}{p{\linewidth}}
\toprule
\textbf{System Prompt} \\
\midrule

\textbf{Instruction:}\\
Analyze the following short text (e.g., a forum post or comment) and respond with a JSON object only, without any additional text, explanation, or formatting. \\

\textbf{Task Objective:}\\
Given the input text:
\begin{verbatim}
---
{text[:4000]}
---
\end{verbatim}

Return exactly one JSON object with the following keys:
\begin{itemize}
    \item \texttt{"sentiment"}: one of \texttt{"positive"}, \texttt{"neutral"}, or \texttt{"negative"}.
    \item \texttt{"dominant\_emotion"}: one of \texttt{"anger"}, \texttt{"joy"}, \texttt{"sadness"}, \texttt{"fear"}, \texttt{"surprise"}, \texttt{"disgust"}, \texttt{"trust"}, \texttt{"anticipation"}, or \texttt{"neutral"}.
    \item \texttt{"is\_conflict"}: a boolean value (\texttt{true} or \texttt{false}). Set to \texttt{true} only if the text contains confrontational, insulting, or offensive language directed toward someone else (e.g., personal attack, hostility, or put-down). Set to \texttt{false} for mere disagreement or negative self-expression.
\end{itemize}

The output must contain only the JSON object and nothing else. \\

\bottomrule
\end{tabular}

\caption{Prompt template for the LLM-based emotion and conflict annotation.}
\label{tab:LLM-judge-emotion}
\end{table}

\begin{table}[ht]
    \small
    \renewcommand{\arraystretch}{0.95}
    \begin{tabular}{p{\linewidth}}
    \toprule
    \textbf{System Prompt} \\
    \midrule

    You are evaluating whether online community posts follow a community-specific writing norm --- a structured format unique to that community, beyond general writing habits.
    Use these label criteria:
    \begin{itemize}
        \item \textbf{yes}: the posts clearly and consistently follow a specific structural format (e.g.\ a required prefix, a fixed field order, a mandatory template);
        \item \textbf{maybe}: there are some recurring structural patterns but they are partial, inconsistent, or only present in a subset of posts;
        \item \textbf{no}: the posts show no detectable community-specific structural format.
    \end{itemize}

    Always respond with a valid JSON object containing exactly:
    \texttt{"label"} (one of: \texttt{"yes"}, \texttt{"no"}, \texttt{"maybe"}),
    \texttt{"norm\_summary"} (if yes/maybe: an actionable checklist of required elements; empty string if no),
    \texttt{"reason"} (explanation; always required).\\

    \midrule
    \textbf{User Prompt} \\
    \midrule

    \begin{minipage}{\linewidth}
    \textit{[Norm Definition]}\\
    A community writing norm is a structured format that members of a specific online community consistently use when posting --- going beyond general internet or AI writing habits to reflect a convention unique to that community.
    Examples: a marketplace community where all posts follow \texttt{Item | Price | Condition}; a bug-report community where posts always include \texttt{Steps / Expected / Actual}; or a community where all posts begin with a required tag like \texttt{[News]} or \texttt{[AMA]}.

    \medskip
    \textit{[Posts]}\\
    The following posts are from the same topic cluster within a community:
    \begin{verbatim}
    1. {post_1}
    2. {post_2}
    ...
    10. {post_10}
    \end{verbatim}

    Do these posts share a community-specific writing norm?

    \medskip
    Respond with a JSON object using exactly these fields:
    \begin{verbatim}
    {
      "label": "<yes | no | maybe>",
      "norm_summary": "<checklist, e.g. 'Posts must include:
                        (1) X, (2) Y'>; empty if no",
      "reason": "<explanation>"
    }
    \end{verbatim}
    \end{minipage}

    \\
    \bottomrule
    \end{tabular}
    \caption{Prompt used for LLM norm judgment of post clusters.}
    \label{tab:LLM-judge-temp-sec4}
\end{table}

\end{document}